\documentclass[prl,reprint,showpacs,superscriptaddress,amsmath,amssymb]{revtex4-1}
\usepackage{dcolumn}
\usepackage{amsmath,amsfonts}
\usepackage{xcolor}
\usepackage{dcolumn}
\usepackage{bm}

\usepackage[utf8]{inputenc}
\usepackage[T1]{fontenc}
\usepackage{mathptmx}
\usepackage{float}
\usepackage{amssymb}
\usepackage{amsmath}
\usepackage{bbm}
\usepackage{graphicx}
\usepackage{color}
\usepackage{ulem}
\usepackage{epstopdf}
\usepackage{textcomp}
\renewcommand{\emph}{\textit}

\begin{document}
\title{Local electrodynamics of a disordered conductor model system measured with a microwave impedance microscope}

\author{Holger~Thierschmann}
\email[]{h.r.thierschmann@tudelft.nl}
\affiliation{Kavli Institute of NanoScience, Delft University of Technology, 
	Lorentzweg 1, 2628 CJ Delft, The Netherlands}
\author{Hale Cetinay}
\affiliation{Faculty of Electrical Engineering, Delft University of Technology, 
	Delft, The Netherlands}
\affiliation{Institute of Environmental Sciences, Faculty of Science, Leiden University, Einsteinweg 2, 2333 CC Leiden, The Netherlands}
\author{Matvey Finkel}
\affiliation{Kavli Institute of NanoScience, Delft University of Technology, 
	Lorentzweg 1, 2628 CJ Delft, The Netherlands}
\author{Allard J. Katan}
\affiliation{Kavli Institute of NanoScience, Delft University of Technology, 
	Lorentzweg 1, 2628 CJ Delft, The Netherlands}
\author{Marc P. Westig}
\affiliation{Kavli Institute of NanoScience, Delft University of Technology, 
	Lorentzweg 1, 2628 CJ Delft, The Netherlands}
\author{Piet Van Mieghem}
\affiliation{Faculty of Electrical Engineering, Delft University of Technology, 
	Delft, The Netherlands}
\author{Teun M. Klapwijk}
\affiliation{Kavli Institute of NanoScience, Delft University of Technology, 
Lorentzweg 1, 2628 CJ Delft, The Netherlands}
\affiliation{Physics Department, Moscow State Pedagogical University, Moscow 119991, Russia.}

\date{\today}


\begin{abstract} 
	We study the electrodynamic impedance of percolating conductors with a pre-defined network topology using a scanning microwave impedance microscope (sMIM) at GHz frequencies. For a given percolation number we observe strong spatial variations across a sample which correlate with the connected regions (clusters) in the network when the resistivity is low such as in Aluminum. For the more resistive material NbTiN the impedance becomes dominated by the local structure of the percolating network (connectivity). The results can qualitatively be understood and reproduced with a network current spreading model based on the pseudo-inverse Laplacian of the underlying network graph. 
     
\end{abstract}
\maketitle

A large class of challenging problems in condensed matter physics emerges from spatially inhomogeneous, co-existing electronic phases and percolation phenomena \cite{basov2011electrodynamics,kirkpatrick1973percolation,van1992theory}. These include superconductor-insulator transitions \cite{gantmakher2010superconductor}, phase transitions in strongly correlated materials \cite{qazilbash2007mott} and many properties observed in quantum materials \cite{keimer2017physics,soumyanarayanan2016emergent}. 
In order to better address these problems, it is advantageous to have experimental access to the electronic properties on a local scale. Since conventional transport experiments measure the electrical properties on a global scale, scanning near field imaging techniques have emerged as key experimental tools \cite{rosner2002high,basov2011electrodynamics,huber2008terahertz,lai2010mesoscopic,ma2015unexpected,gramse2017nondestructive,atkin2012nano,anlage2007principles}. These techniques use high frequency signals that are scattered off or are reflected from a sharp metallic probe, and provide quantitative information about the electric and dielectric properties of the sample material in the vicinity of the probe tip. 
In this manner the electrodynamic response can be studied even for electrically disconnected, small conductive patches in an insulating environment, without the need for external electrical contacts and a fully conducting path through the sample \cite{lai2010mesoscopic,qazilbash2007mott,huber2008terahertz,ma2015mobile,de2016spatial,gramse2017nondestructive}.
Significant progress has been made towards a quantitative interpretation of the signal \cite{anlage2007principles,lai2008modeling,huber2012calibrated,gramse2014calibrated,buchter2018scanning}.
 \begin{figure}
	\centering
	\includegraphics[width=1\linewidth]{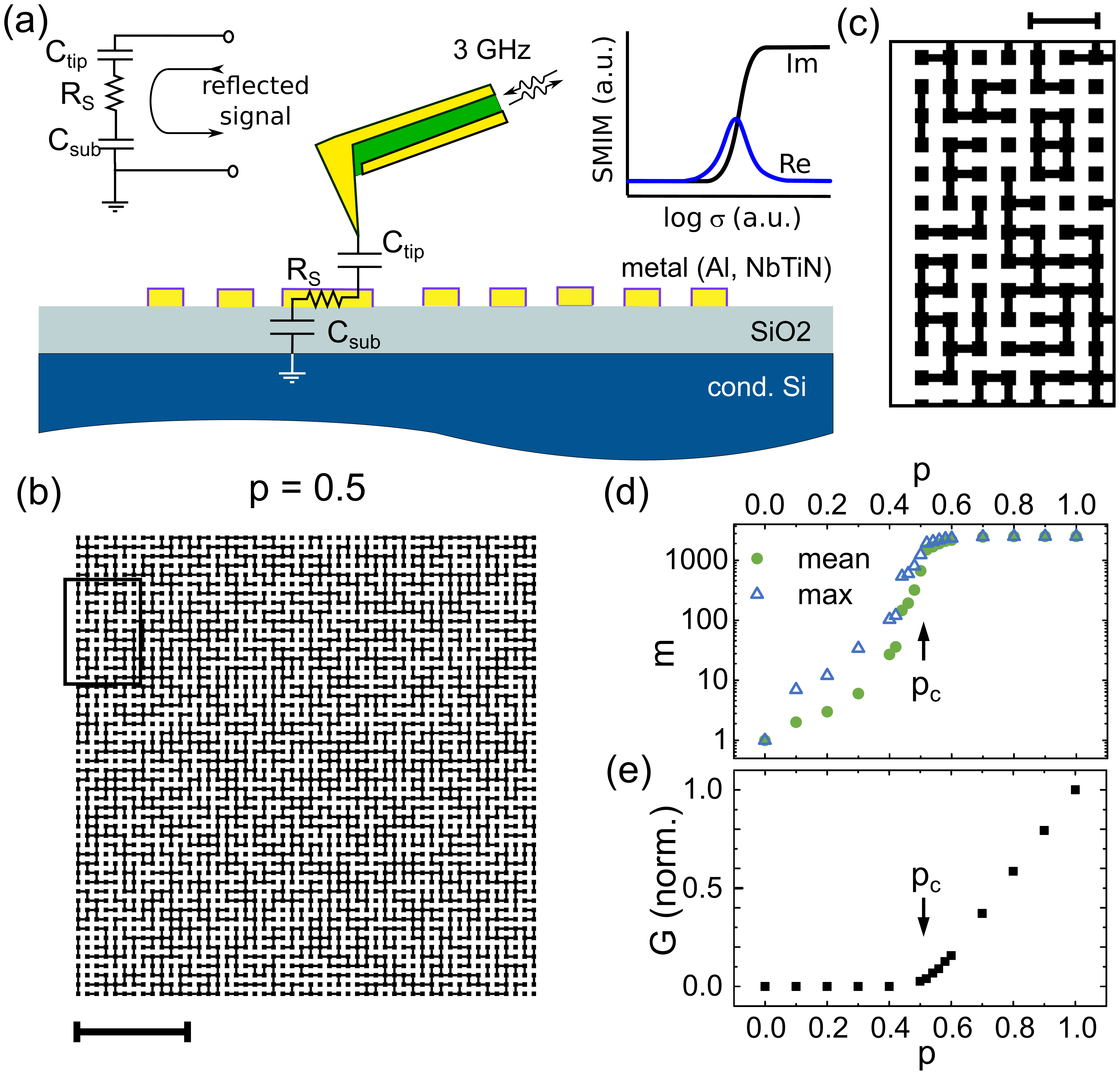}
	\caption{\textbf{(a)} Scanning Microwave Impedance Microscopy (sMIM): a 3GHz signal is launched to a transmission line cantilever and is reflected due to near field interactions at the tip. The near field interactions can be described in a lumped element circuit (left inset) consisting of capacitors $C_\text{tip}$ and $C_\text{sub}$ and a resistor $R_s$. The characteristic sMIM response curve for the imaginary (sMIM-Im) and real (sMIM-Re) part of the reflected signal are coupled through the sample conductivity $\sigma$ (right inset). \textbf{(b)} Bond-percolated network pattern for a disordered conductor model system with percolation number $p$~=~0.5. Scale bar: 5 $\mu m$. \textbf{(c)} Close up of the region of the network indicated with a frame in b). Scale bar: 1 $\mu$m. \textbf{(d)}  Mean cluster mass $m_\text{mean}$ and mass of the largest cluster $m_\text{max}$ calculated for each network pattern versus $p$. \textbf{(e)}  Calculated dc conductance $G$ between the left and the right edge of the networks versus $p$, normalized to $G(p=1)$. $p_c$ indicates the IMT.
	}
	\label{fig:1}
\end{figure}
However, an important question remains as to how short and long ranged structural correlations of different electronic phases, arising inherently in electrically inhomogeneous materials, modify the electrodynamic response. Studying such contributions from disorder in an experiment is difficult because the spatial details during a phase transition typically follow seemingly random distributions and thus can not be known \textit{a priori}.

Here we address through experiments and calculations the local impedance of conductors that exhibit a precisely known spatial distribution of disorder. 
We have designed and realized a set of microscopic two-dimensional percolated networks from different metallic materials using lithography techniques. The networks serve as model systems for a disordered conductor that exhibits an insulator-to-metal transition (IMT). 
In order to measure the electrodynamic impedance locally, we use a room temperature scanning microwave impedance microscope (sMIM). We find that the impedance is strongly affected by the local topology of the disordered network as well as by the resistivity of the material.

The sMIM used for our experiments is based on an Asylum Cypher Atomic Force Microscope with a PrimeNano Scanwave extension.
sMIM uses a shielded cantilever as depicted in Fig.~\ref{fig:1}(a) to guide a microwave tone to the metallic tip were the signal is reflected back into the cantilever transmission line and fed into a microwave readout circuit \cite{lai2010mesoscopic,lai2008modeling}.
When the tip (typical apex diameter $\approx 100 $ nm) is far away from the sample, an impedance matching circuit and a common mode cancellation loop suppress signal reflection. As the tip is landed and scanned over a sample, the local material properties in the vicinity of the probe modify the generally complex numbered tip impedance $Z$ (Fig.~\ref{fig:1}(a)), giving rise to changes in the real and imaginary component of the reflected microwave signal. 
The disordered conductor model systems are patterned on a 100 nm thick dielectric SiO$_2$ layer (with relative permittivity $\epsilon_r = 3.9$) that covers a conductive Si substrate. The impedance $Z$ is therefore described in a lumped element circuit in which the substrate acts as a ground plane. The circuit consists of a series network of two capacitors $C_\text{tip}$ and $C_\text{sub}$, representing the coupling between the tip and the sample, and the sample and the substrate, respectively. A resistor $R_\text{s}$ takes into account resistive losses inside the network. Hence,

\begin{eqnarray}
Z =  R_\text{s}  + \frac{1}{i2\pi f} (\frac{1}{C_\text{tip}}+\frac{1}{C_\text{sub}}),
\label{eq:ZsMIM}
\end{eqnarray}

with $f$ being the applied frequency (3 GHz). The real and imaginary component of the reflected signal are demodulated in the microwave readout circuit into two dc-signals, which correspond to the resistive (sMIM-Re) and the capacitive (sMIM-Im) component of the tip-sample admittance $Y = Z^{-1}$.
The sMIM signal allows for an analysis of material properties because $R_\text{s}$, $C_\text{tip}$ and $C_\text{sub}$ are affected on a microscopic level by the electrical conductivity $\sigma$ of the conductive patch underneath the tip. Therefore, the signals of sMIM-Im and sMIM-Re are coupled. When $\sigma$ changes homogeneously, one obtains characteristic curves for sMIM-Im and sMIM-Re as sketched in the right inset in Fig.~\ref{fig:1}(a). sMIM-Im exhibits a transition from low to high signal as $\sigma$ increases, while sMIM-Re approaches 0 at both extrema and exhibits a peak in between.

We are interested in the sMIM response for an inhomogeneous sample. Therefore, we have designed a conductor model system consisting of a set of nano-structured metallic two-dimensional bond-percolated networks. We aim to probe the impedance with spatial resolution higher than the scale of disorder and to study spatial variations across the sample as a result of the disordered electrodynamic landscape. 
Therefore, we design each network to consist of a $50 \times 50$ square grid of nominally $200$ nm$ \times 200$ nm sized, metallic squares which we will refer to as \textit{sites}. In comparison, the apex diameter of the sMIM tip is of the order of 100 nm. Hence, the sMIM signal can be assigned to the particular site underneath the tip (see SI). Each site can be connected with its 4 nearest neighbours by $100$ nm $\times 200 $ nm metallic \textit{bonds}. Disorder is introduced by randomly placing only a fraction of bonds in the network. In this fashion, the IMT is modeled by a series of patterns for which the number of bonds increases. Each pattern is then characterized by the total fraction of bonds $p$.  
The fully insulating state corresponds to an entirely disconnected network with no bonds, $p=0$. For the fully metallic state $p=1$.
When $0<p<1$, clusters of connected sites form, giving rise to a highly disordered pattern as displayed for the case $p=0.5$ in Fig.\ref{fig:1}(b) and in the closeup in Fig.~\ref{fig:1}(c). 
Each cluster can be characterized by its mass $m$, given by the number of connected sites in that cluster.
For small $p$ the clusters consist of only a few sites, hence the mean $m$ as well as the largest cluster $m_\text{max}$ for a given $p$ are small [cf. Fig.~\ref{fig:1}(e)]. As $p$ is increased the clusters grow and eventually merge with their neighbors. 
The IMT coincides with the emergence of the so-called spanning cluster at a critical percolation number $p_c$. Here, $m_\text{max}$ is sufficiently large to establish a continuous conductive path across the whole network, rendering the previously insulating system a conductor at the global scale. This is directly reflected in the conductance $G$, as calculated for each $p$ \cite{van2017pseudoinverse} between the left and the right edge (see SI for details), which becomes non-zero at $p_c = 0.5$, as expected from percolation theory for a 2D square network \cite{stauffer1979scaling}. (The complete set of patterns is shown in the SI.)

We have fabricated the set of patterns from two different materials with different resistivity. It allows us to study the local impedance for a wider range of parameters. As a low resistivity material we choose a 25 nm aluminum (Al) thin film with a sheet resistance $R_\text{sheet}^\text{Al}=1.04$ $\Omega/\square$, patterned with standard lift-off techniques. As a high resistivity material we use sputtered, 10 nm thick NbTiN \cite{thoen2016superconducting}, with $R_\text{sheet}^\text{NbTiN} = 10.2~\text{k}\Omega/\square$. 

\begin{figure}
	\centering
	\includegraphics[width=1\linewidth]{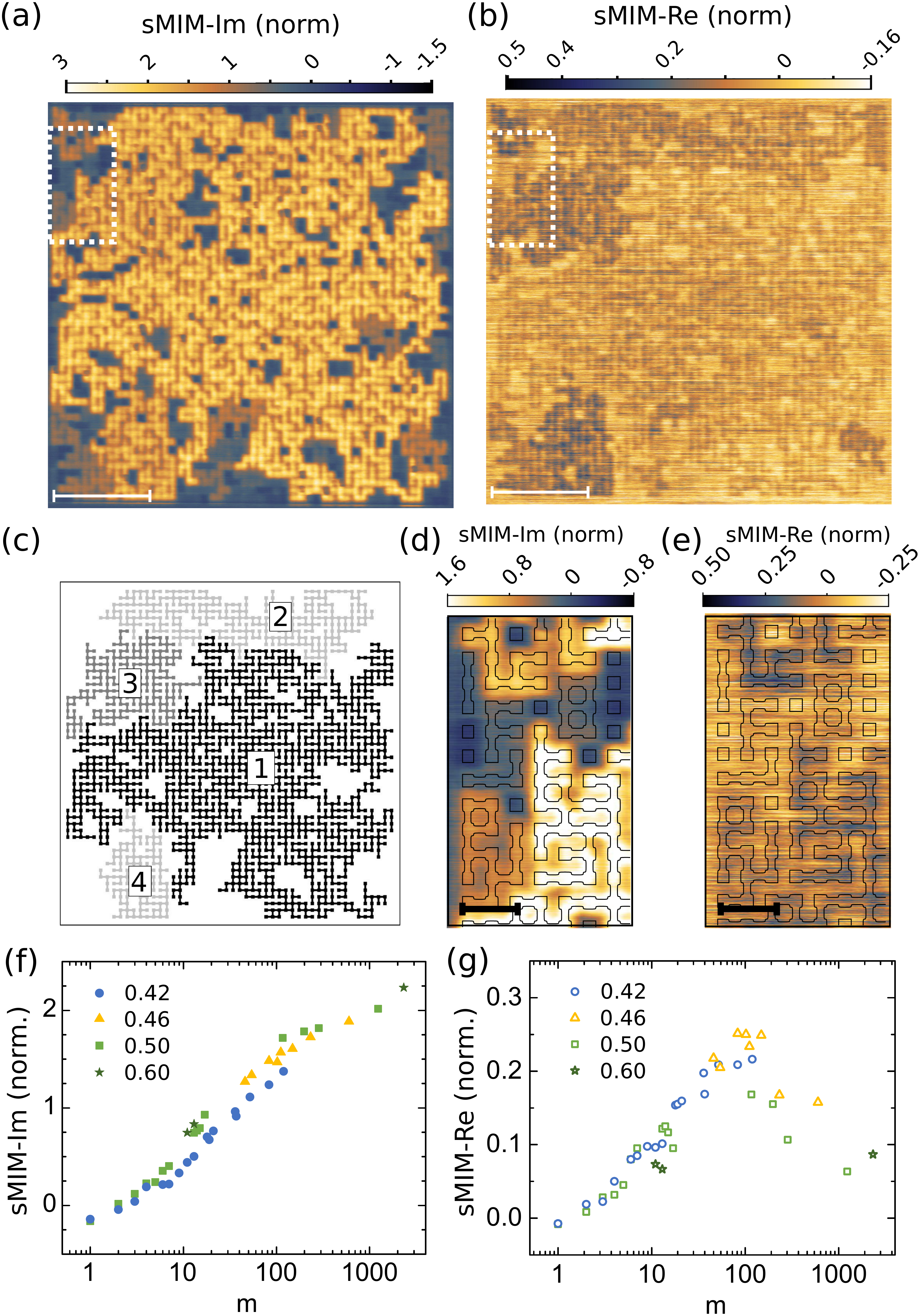}
	\caption{\textbf{(a)} sMIM-Im image and \textbf{(b)} sMIM-Re image of the conductor model system realized from Aluminum with $p=0.5$. The underlying network is identical with the pattern depicted in Fig.\ref{fig:1}(b). Scalebar: 5 $\mu$m. \textbf{(c)}  Network pattern for $p = 0.5$ with all clusters removed except for the 4 largest ones, labelled 1 - 4.  \textbf{(d)} Close up of sMIM-Im and \textbf{(e)} sMIM-Re in the region indicated in (a) and (b) with a white frame. As a guide to the eye, the underlying pattern has been added to the images. Scale bar: 1 $\mu$m. \textbf{(f)} Mean sMIM-Im and \textbf{(g)} mean sMIM-Re signal of various clusters versus cluster mass $m$. The data were obtained from Al samples with $p = 0.42, 0.46, 0.5$ and $ 0.6$.}
	\label{fig:Al}
\end{figure}

Figure \ref{fig:Al} shows typical sMIM images obtained for the network made out of Al  with $p =0.5$. The data have been normalized with respect to a reference sample (see SI) and
they are nulled with respect to the SiO$_2$ dielectric layer. 
Figure \ref{fig:Al}(a) shows the imaginary component sMIM-Im, corresponding to capacitive contributions to the impedance. In Fig.~\ref{fig:Al}(b) we show the real part, sMIM-Re, representing the ohmic contributions. Both images reveal a structure that is different from the underlying network depicted in Fig.~\ref{fig:1}(a). Fig.~\ref{fig:Al} shows that sMIM-Im, i.e. the capacitance, is large (bright) over large areas, while only smaller patches show a small response (dark). sMIM-Re indicates that ohmic resistance contributes to the impedance only in a few, smaller regions.
\begin{figure}
	\centering
	\includegraphics[width=1\linewidth]{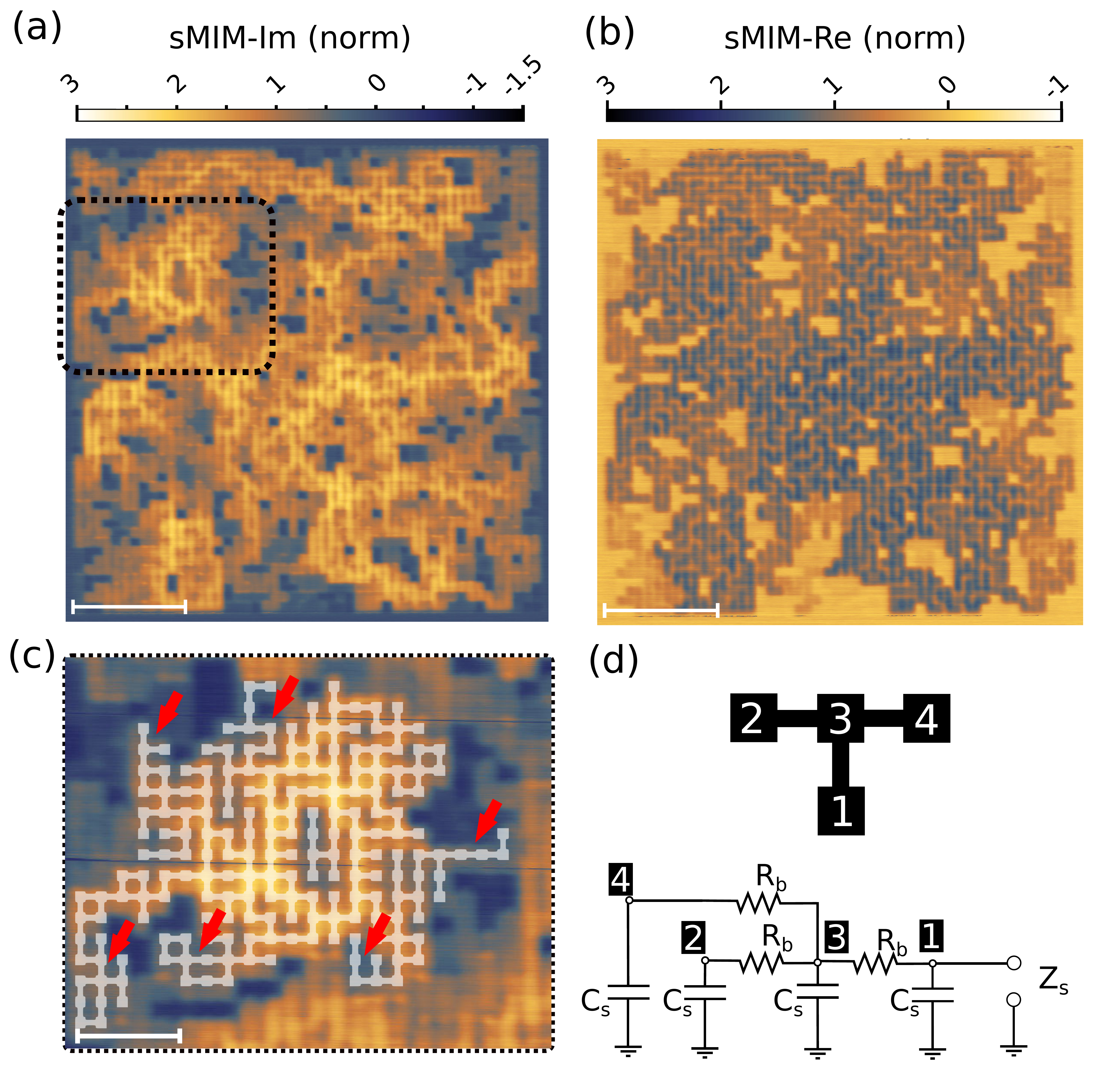}
	\caption{\textbf{(a)} sMIM-Im and \textbf{(b)} sMIM-Re response of the pattern realized from NbTiN with $p=0.5$. Scale bar: 5 $\mu $m. \textbf{(c)} Close-up view of sMIM-Im of cluster 3, indicated with a frame in (a). The geometry of cluster 3 is indicated in gray. The colorscale is the same as in (a). Red arrows indicate polyps of the cluster with low connectivity, yielding low signal. Scale bar: 2~ $\mu$m. \textbf{(d)} Impedance network model used to calculate the sample impedance $Z_s$ for a cluster with $m=4$ with the tip at site 1. $C_s$ denotes the capacitance of a site to substrate ground, $R_b$ corresponds to the resistance of a bond connection.}
	\label{fig:NbTiN}
\end{figure}
Since the pattern was fabricated from a single Al film the observed spatial variations can not be explained with changes in the local conductivity of the Al. We attribute the difference to disorder and thus to the formation of clusters in a network with non-trivial topology.

Figure~\ref{fig:Al} (c) depicts the network underlying the sMIM data (cf. Fig.\ref{fig:1}(a)), however, we have removed all sites and bonds from this pattern, except the clusters with the largest cluster mass $m$, labeled 1 to 4.
The resulting shape emerging from these clusters strongly resembles the image in Fig.~\ref{fig:Al}(a). This suggests a correlation between the capacitive impedance (sMIM-Im) and the cluster mass $m$. We further observe that areas with a strong sMIM-Re signal coincide with clusters 2, 3 and 4. The connection between cluster mass $m$ and impedance appears to hold also for smaller $m$. It can be inferred from the close-up view of sMIM-Im and sMIM-Re in Fig.~\ref{fig:Al}(d) and (e), where we have added the underlying network pattern from Fig.~\ref{fig:1}(c) as a guide to the eye. Within a cluster the sMIM signals appear to be uniform. We can therefore calculate the mean signal obtained for each cluster and assign the result to the corresponding $m$. This is shown in Figs.~\ref{fig:Al}(f) and (g) for an arbitrary set of clusters, obtained from various Al patterns of different $p$. Figure~\ref{fig:Al}(f) reveals a monotonous increase with $m$ for sMIM-Im. The maximum slope occurs approximately around $m=100$. In contrast, sMIM-Re in Fig.~\ref{fig:Al}(g) exhibits a peak here, while it decreases towards 0 for smaller $m$ and approaches a small, but non-zero value towards $m=2500$.

\begin{figure}
	\centering
	\includegraphics[width=1\linewidth]{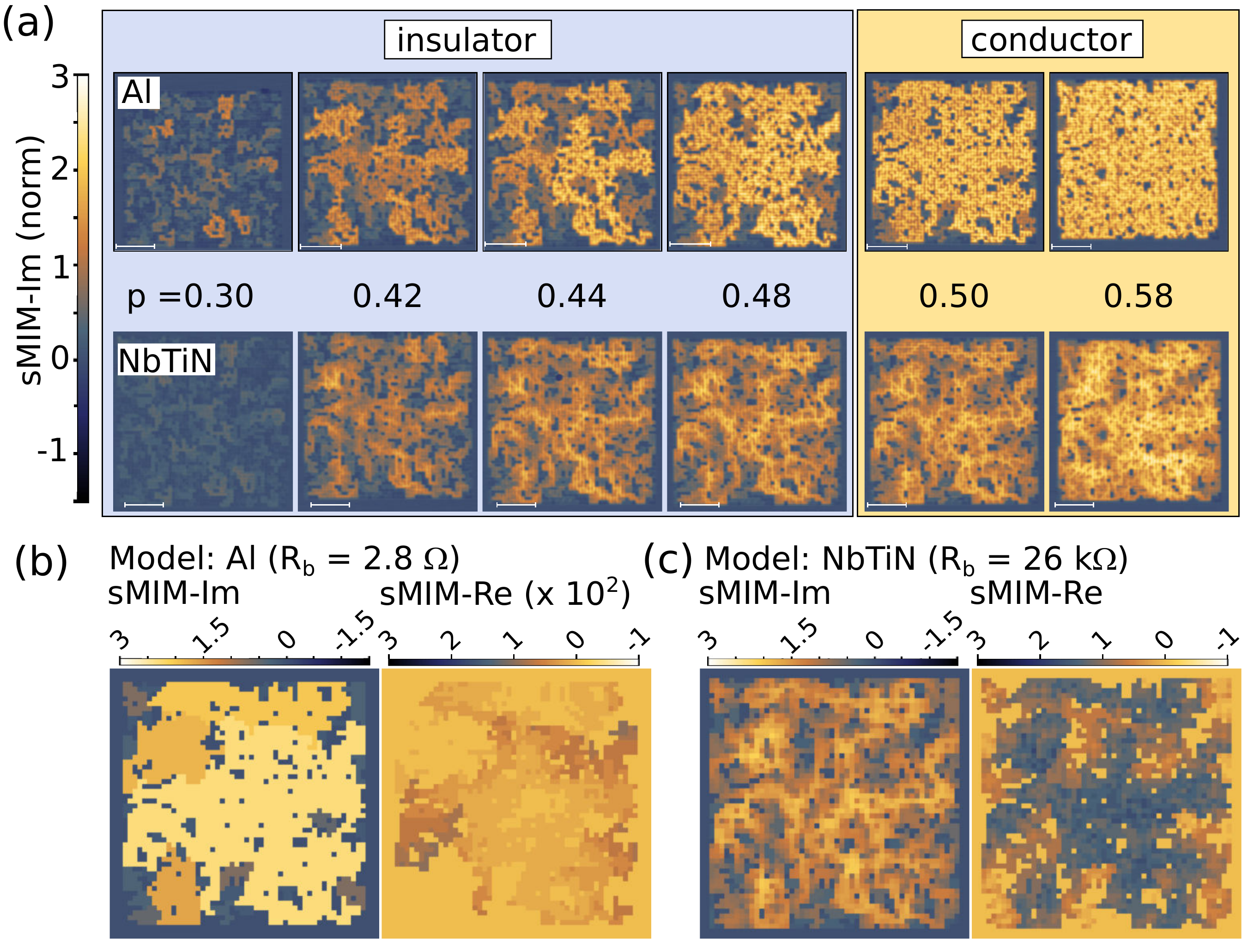}
	\caption{\textbf{(a)} Experimental sMIM-Im images for Al (top row) and NbTiN (bottom row) for various percolation numbers $p$ from the globally insulating (blue background) to the conductive (yellow background) state. \textbf{(b)},\textbf{(c)} Model calculations of the sMIM-Im and sMIM-Re response of the network with $p = 0.5$ for Al and NbTiN, respectively, using an impedance network model. The detailed parameters are given in the SI. Note that the scale of sMIM-Re in (b) is scaled by a factor 10$^2$.  
	} 
	\label{fig:model}
\end{figure}

When the exact same network pattern is made from a less conductive material, the corresponding sMIM images change drastically. This is shown in Fig.~\ref{fig:NbTiN} where the network has been realized from NbTiN. sMIM-Im (Fig.~\ref{fig:NbTiN}(a)) reveals a highly non-uniform distribution of the imaginary impedance that does not bear a clear resemblance with the clusters in Fig.~\ref{fig:Al}(c). Instead, a backbone-type structure becomes visible. sMIM-Re (Fig.~\ref{fig:NbTiN}(b)), in contrast, does not show a backbone structure. Here, the cluster delimitation can be inferred more clearly. However, signal variations within the clusters are significant, which renders an analysis as done for Al in Fig.~\ref{fig:Al} with sMIM as a function of cluster mass $m$, not applicable. In Fig.~\ref{fig:NbTiN}(c) we therefore directly compare sMIM-Im with the underlying network topology for the region of cluster~3 (cf. Fig.~\ref{fig:Al} c). We observe that the capacitance (sMIM-Im) is small at the outside branches and polyps of the cluster. It becomes large towards the inside, in particular in a well connected, ring-shaped region, circling the cluster's center.

sMIM-Im images of the IMT around $p_c$ are displayed in Fig.~\ref{fig:model}(a), for Al (top row) and NbTiN (bottom row), highlighting the evolution of the local impedance as structural correlations and network topology evolve. 

We can understand the observations in the local impedance within the lumped element picture introduced in Fig.~\ref{fig:1}(a), taking into account the electrodynamic environment of the scanning probe experiment and the detailed disordered network pattern. We replace the resistor $R_s$ and the capacitor $C_\text{sub}$ in Eq.~\ref{eq:ZsMIM} by an impedance network that reflects the network topology and the capacitive coupling at each site to the substrate ground plane.
Figure~\ref{fig:NbTiN}(d) displays such a network for a small cluster with $m = 4$, with the probe tip positioned at site 1. Each bond connection is represented by a resistor $R_\text{b}$. Each site is taken into account by a capacitor $C_{s}$ to the substrate.
 
When $R_\text{b}$ is small compared to the impedance of a single site to ground $Z_{Cs} =(i2\pi f C_s)^{-1}$, the microwave currents injected at the probe tip can spread easily across the whole cluster. This applies for Al, where $R_\text{b} = 2.8$ $\Omega$, while $Z_{Cs} \approx 1.4$ M$\Omega \gg R_\text{b}$ (with $C_\text{s} = 4\times 10^{-2}$ fF, see SI). As a result, all site capacitors $C_s$ of a cluster contribute approximately equally to the total capacitance $C_\text{sub}$, such that $C_\text{sub} \approx m C_s $. This explains the absence of variations within a cluster as well as the observed scaling of the capacitive impedance, sMIM-Im, with $m$.
However, when $R_\text{b}$ becomes sufficiently large, as is the case for NbTiN ($R_\text{b} = 26~k\Omega$), current spreading is impeded. Capacitive contributions to the total impedance, therefore, get suppressed when the sites are connected only through a high resistance current path, for instance via several bonds in series. This is the case at the polyps and dead ends in cluster 3 in Fig.\ref{fig:NbTiN}(c) (arrows). Here the impedance is large due to a small total capacitance $C_\text{sub}$. In contrast, in regions where the local connectivity is high, for instance towards the center ring of cluster 3, current injected at the tip can spread over a large number of sites through many parallel current paths. This leads to a smaller impedance due to a large capacitance, and thus to a high sMIM-Im signal. In this picture, the impedance at a particular site reflects the local connectivity of the network, giving rise to the observed backbone type structure.

The RC-network picture indicates how the capacitance to the substrate ground plane $C_\text{sub}$ may serve as a tuning knob for the range of contrast in a sMIM experiment. Modifying thickness and permittivity of the dielectric will shift the quantitatively sensitive sMIM response to smaller or larger patches of film, i.e. to different scales of disorder. 

Based on the RC-network approach, we use a network current spreading model and the pseudo-inverse Laplacian of the underlying network graph \cite{van2017pseudoinverse} to compute the impedance $Z$ at each site and, using Eq.\ref{eq:ZsMIM}, to compare it with the experimental data (see SI). The results are shown in Fig.~\ref{fig:model} (a) and (b) for Al and NbTiN, respectively. The computed sMIM-Im images reproduce the experimental data (cf. Fig.~\ref{fig:Al}(a) and Fig.~\ref{fig:NbTiN}(a))) very well. The computed sMIM-Re signal (cf. Fig.~\ref{fig:Al}(b) and Fig.\ref{fig:NbTiN}(b)), however, deviates more from the experiments. Interestingly, while for NbTiN deviations are small and mainly occur in samples with larger $p$ (see SI), deviations are particularly strong for Al, where the sMIM-Re signal magnitude, compared to the corresponding sMIM-Im, is orders of magnitude smaller than in the experiments. The differences between the model prediction and the data on Al therefore suggest that the simple RC network model used here is incomplete.

The present experiments highlight the electrodynamic environment of the tip provided by the inhomogeneous sample. It is reminiscent of the importance of the electrodynamic environment in scanning tunneling experiments, such as in a Josephson current based tunneling experiment (for example Ast et al. \cite{ast2016sensing}). Here, the metallic tip of the scanning probe acts as a resonant antenna interacting with the Josephson oscillations. 
Our experiments indicate that for electrically inhomogeneous materials, such as a structured film of Al or disordered superconductors such as NbTiN, TiN or InO \cite{sacepe2008disorder,dubouchet2019collective} the sample under study also contributes to the effective electrodynamic environment.

Finally, we note that it may be an interesting route in the future to explore techniques to invert the procedure described here, i.e. to construct the underlying disordered network in an unknown material from a sMIM image.     

In conclusion, we have studied experimentally the role of disorder for the local impedance in a percolating conductor model system using microwave impedance microscopy. We find that structural properties such as cluster size, network topology and local connectivity can significantly influence the local electrodynamic environment, depending on the material resistivity and the degree of disorder in the system.

\begin{acknowledgements}
We like to thank David J. Thoen for help with the NbTiN sample fabrication.
We acknowledge funding through the European Research Council Advanced Grant No.~339306 (METIQUM).

\end{acknowledgements}


%

\newpage
\renewcommand{\thefigure}{S\arabic{figure}}
\setcounter{figure}{0}
\mbox{}
\thispagestyle{empty}
\newpage

\onecolumngrid

\section{Supplementary Information}

\subsection{sMIM reference sample}
\begin{figure}[h]
	\includegraphics[width = 0.7 \linewidth]{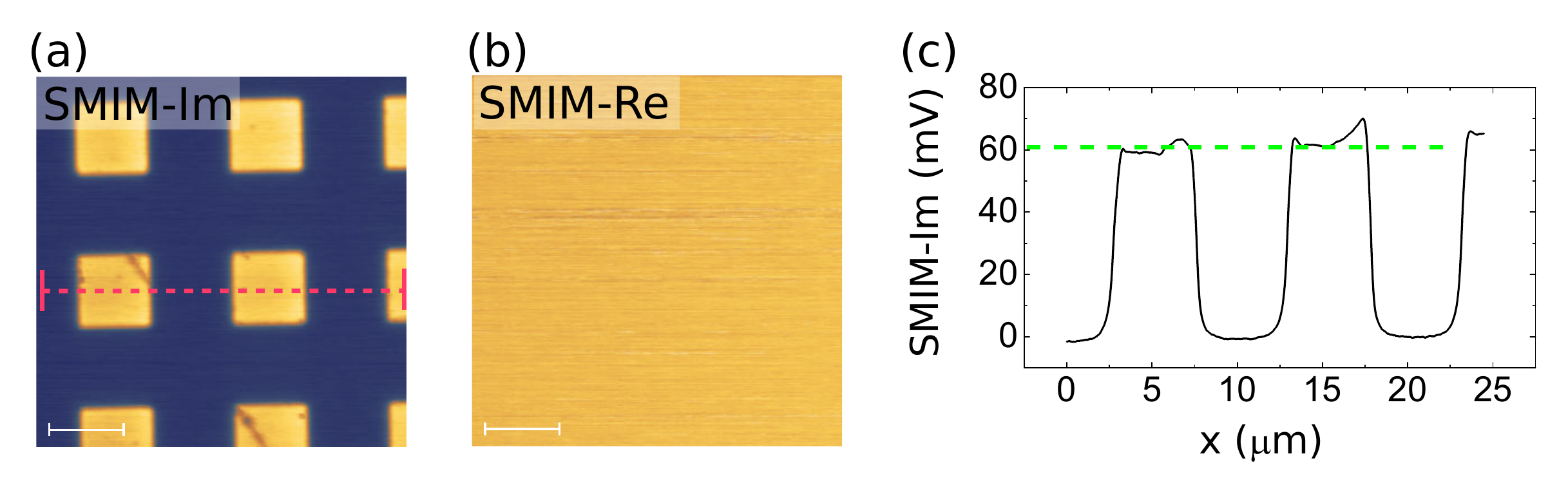}
	\caption{\textbf{(a)} sMIM-Im and \textbf{(b)} sMIM-Re from a measurement of the calibration sample. Since the sample consists of highly conductive aluminum on a dielectric substrate, contrast is only visible in sMIM-Im while ay cnontrast is absent in sMIM-Re. The scale bar corresponds to 5 $\mu$m. \textbf{(c)} Cross section taken from (a). The sMIM-Im amplitude obtained for the Al squares (green dashed line) from frequently repeated scans of the reference sample is taken as a reference for the sMIM measurements shown in the main text.}
	\label{fig:calib}
\end{figure}

In order to be able to directly compare sMIM data obtained from different structures, we use as a reference sample consisting of $5 ~\mu$m $\times 5~\mu$m Al squares on $SiO_2$, as provided by the sMIM manufacturer PrimeNano Inc (Fig.\ref{fig:calib}). While collecting the data from the NbTiN and Al networks, we frequently scanned the reference sample. This allows us to normalize the sMIM data for each network shown in the main text, with respect to the mean contrast in the sMIM-Im channel obtained from the reference sample. In this manner the sMIM data for different structures become comparable and signal changes due to tip wear-off of electronic drift get removed.

\newpage

\subsection{Topography characterization of networks}
\begin{figure}[h]
	\includegraphics[width = 0.7\linewidth]{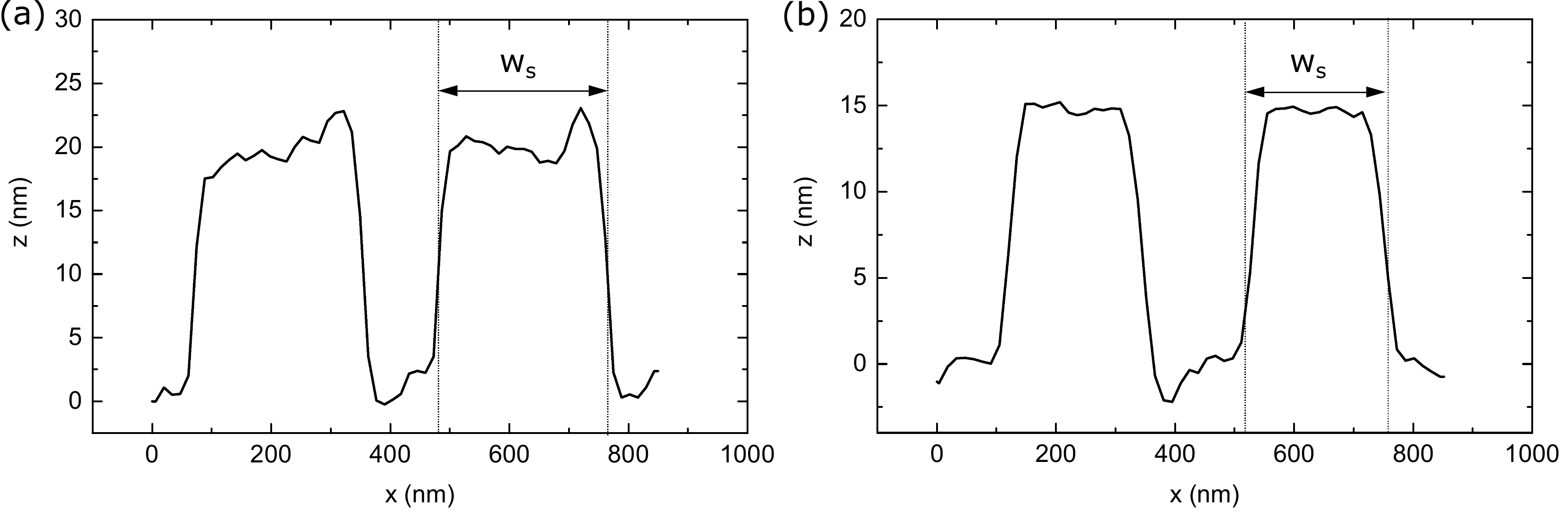}
	\caption{Topography scans of two adjacent disconnected sites. \textbf{(a)} Aluminum. \textbf{(b)} NbTiN. The size of the sites is denoted with $w_s$. Note that these traces were taken with a different tip than the one used for the sMIM measurements.}
	\label{fig:profiles}
\end{figure}

Figure~\ref{fig:profiles} (a) and (b) depict topography line scans of two disconnected sites fabricated out of Aluminum and NbTiN, respectively. Note that these scans were taken with a different, sharper tip than the one used for the sMIM measurements. While the size of the NbTiN sites ($w_s \approx 230$ nm) is close to the nominal value (200 nm), the Al site is larger ($w_s \approx 260$ nm). This is a result of the different fabrication techniques, which consisted of a lift-off process for Al, and a dry-etch step for the NbTiN sample.  

\newpage
\subsection{Contributions on the sMIM signal from neighboring sites}
\begin{figure}[h]
	\includegraphics[width=0.9\linewidth]{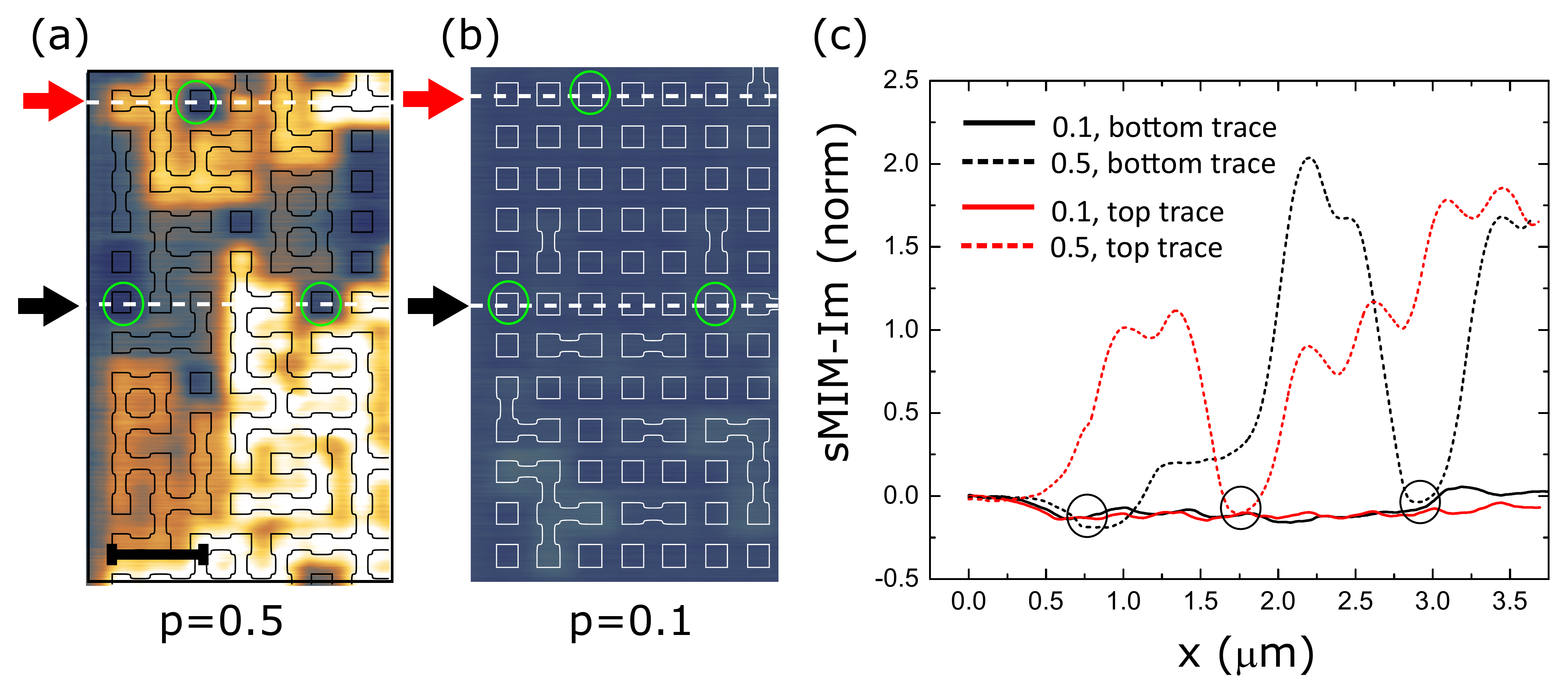}
	\caption{\textbf{(a)} sMIM-Im for the Al sample with p=0.5. Close up of the same region as shown in Fig.~\ref{fig:Al}(d) in the main text. The circles indicate individual singlets that are surrounded by larger cluster. \textbf{(b)} sMIM-Im image of the same set of sites as in (a) for $p = 0.1$. Circles indicate the same sites as in (a), which are now surrounded by other singlets and small clusters. \textbf{(c)} Horizontal traces extracted from (a) (dashed lines) and (b) (solid lines), along the dotted lines indicated with a red (top) and black (bottom) arrow. Circles indicate the sMIM-Im signal at those locations marked with cirles in (a) and (b). We find that despite strong differences in the environment of these sites, the difference in the sMIM-Im signal are barely visible. This confirms that the measured signal originates from the microwvae current being injected only at a single site underneath the tip, justifying the current spreading model used for the calculations.}
\end{figure}

\newpage
\subsection{Tip shape estimate}
\begin{figure}[h]
	\includegraphics[width =1\linewidth]{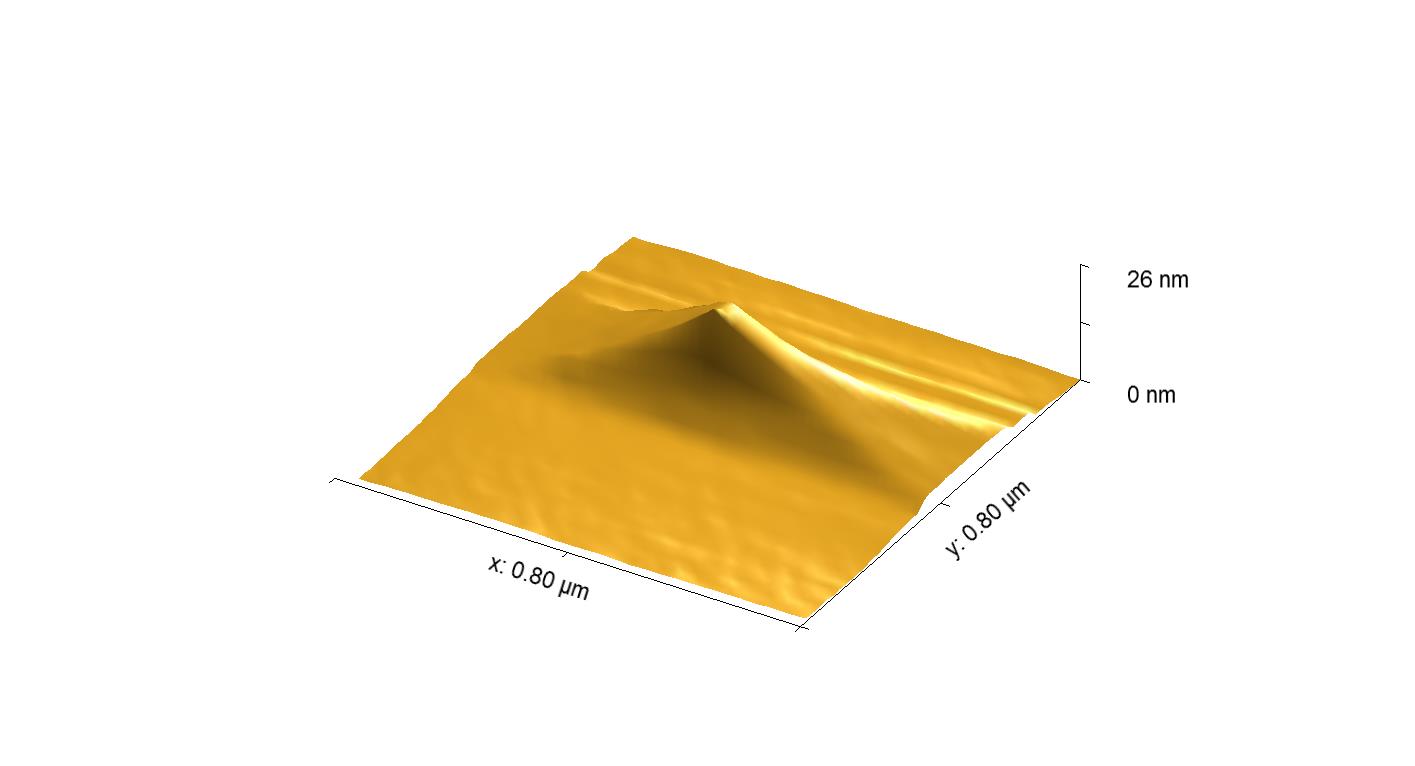}
	\caption{ We have used the \textit{Gwyddion} blind tip shape estimate algorithm \cite{nevcas2012gwyddion} based on the topography images to extract a quantitative estimate of the tip geometry.}
	\label{Sfig:tip}
\end{figure}

\newpage
\subsection{Disordered conductor model systems}
\begin{figure}[h]
	\includegraphics[width = \linewidth]{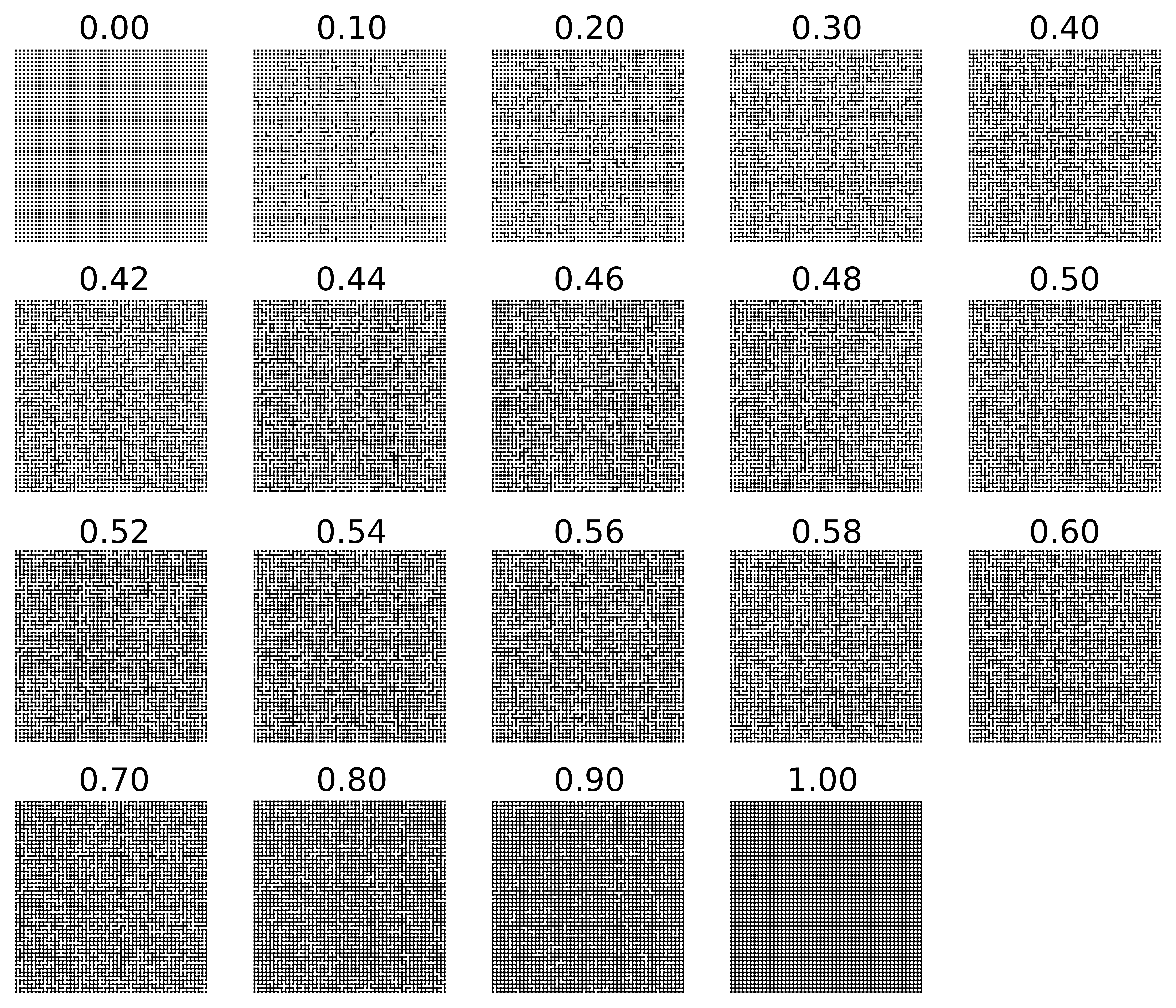}
		\caption{Percolated network patterns used to realize the model systems for the insulator-metal-transition. The patterns consist of $50 \times 50$ sites (total number N = 2500). The numbers indicate the percolation number $p$, i.e. the fraction of occupied bonds (total number $N_b = 4900$) in the network. The patterns have been designed to be incremental, i.e. when increasing $p$ for example from 0.1 to 0.2, the bond configuration for $p$~=~0.1 is maintained and 10 \% more bonds are added to the system.}
\label{Sfig:patterns}
\end{figure}
\newpage

\subsection{Calculation of model parameters $C_s$, $C_\text{tip}$ and $R_b$}
The capacitance between a single site and the substrate ground plane is calculated using a textbook closed-form expression for a microstrip line \cite{pozar2009microwave} (corresponding to metallic plate above an infinite ground plane), with the goal to use a simple approximation that also takes into account stray fields.
\begin{equation}
	C_s = \frac{\epsilon_r w_s [w_s/h + 1.393+0.667 \text{ln}(w_s/h + 1.444)]}{120 \pi c},
\end{equation}
 with site dimensions $w_s=250$ nm (estimated from the topography images), dielectric thickness $h = 100$ nm, the dielectric constant $\epsilon_r = 3.9$ and the vacuum speed of light $c$.
This yields $C_s = 4 \times 10^{-2} $ fF.

The tip-site capacitance $C_\text{tip}$ is calculated in a similar fashion. From the topography images obtained along with the sMIM data, we infer that the sMIM tip is strongly worn down (cf. blind tip estimate in Fig. \ref{Sfig:tip}). We empirically find good agreement for $C_\text{tip} = 4 $ fF, corresponding, in a parallel plate capacitor model, to a tip size of 200 nm interacting capacitively with the sample, and 1 nm thin layer of natural $Al_2O_3$  covering the Al structures ($\epsilon_r = 10$).

Using this value to model the experiments on NbTiN also yields good agreement. This is a surprising since obviously these structures are not expected to be exhibit an $Al_2O_3$ layer. However, while little is known about the oxidation of NbTiN structures under ambient conditions, some studies suggest that a dielectric film is formed \cite{jouve1996xps}. In our case, even though dielectric constant and thickness of such a film are likely to be different compared to the Al samples, apparently the resulting $C_\text{tip}$ is of similar order. 

The resistance of each bond, $R_b$ is calculated using the measured sheet resistance ( $R_\text{sheet}^\text{Al}=1.04$ $\Omega/\square$, $R_\text{sheet}^\text{NbTiN} = 10.2$ k$\Omega/\square$) and the geometry of the bond connection $w_b = 100$ nm and $l_b = 200$ nm. Including the resistance of half a site on each side of the bond connection adds $R_\text{sheet}$ to the bond resistance, $R_b = R_\text{sheet} \frac{l_b}{w_b} + R_\text{sheet}$. This yields $R_b^\text{Al} = 2.5 \Omega$ and $R_b^\text{NbTiN} = 26 \Omega$, as used in the main text.

\subsection{Impedance network model}

In order to calculate the electrical properties of our disordered conductor model systems we follow an approach described by Van Mieghem et al. in Ref. \cite{van2017pseudoinverse}. We convert each of the percolated conductor patterns shown in fig. \ref{Sfig:patterns} into a matrix $P$ and construct the corresponding weighted adjacency matrix $\tilde{A}$, yielding an $N\times N$ matrix ($N=2500$ being the total number of sites) which only has non-zero entries $q_{ij} = 1/R_b$ if the sites $i$ and $j$ are directly connected through a single bond in $P$. The capacitive connection to the substrate ground plane in the experiments is taken into account by adding a $(N+1)^{th}$ ground node to $\tilde{A}$ that is directly connected to every other site via the complex impedance $(i2\pi f C_s)^{-1}$. This yields the modified adjacency matrix $\tilde{A}_\text{gnd}$
from which we can construct the Laplacian $\widetilde{Q}$
\begin{equation}
\widetilde{Q} = \text{diag} \left({\sum\limits_{k=1}^{N+1} \tilde{A}_\text{gnd,ik}}\right) - \tilde{A}_\text{gnd}.
\end{equation}

The complex effective impedance between the two nodes $a$ and $b$ is then given by \cite{van2017pseudoinverse}
\begin{equation}
Zs_{ab} = Q^\dagger_{aa} + Q^\dagger_{bb} - 2Q^\dagger_{ab}
\end{equation}
where $Q^\dagger$ is the pseudo-inverse of the weighted Laplacian $\tilde{Q}$. 

The local impedance $Z_s$ of the network at each site $i$, as relevant in our experiment, is then obtained by calculating the impedance between site $i$ and the ground node, $Zs_{i,gnd}$.

In addition to $Zs_{i,gnd}$, signal reflection at the cantilever tip is also determined by the capacitive tip-site coupling, $C_{tip}$. Hence,

\begin{equation}
Z_i = Zs_{i,gnd} + \frac{1}{i (2\pi f) C_{tip}}.
\end{equation}

The dissipative and capacitive component of the measured sMIM signal as plotted in Fig. 4 (a) and (b) in the main text, sMIM-Im and sMIM-Re, are then obtained by taking the real and imaginary components of the total admittance $Y_i = Z_i^{-1}$ for each site $i$ in the network.

The macroscopic conductance $G$ between the left and the right edge of each pattern as plotted in Fig.\ref{fig:1}(d) in the main text, is obtained by summing up the conductance between each site on the left edge and all the sites of the right edge. $G$ is then given by the real component of the result. Since $G$ corresponds to a dc conductance, we ensure that contributions from ac currents through the ground node are suppressed in this calculation by letting $C_s \rightarrow 0$.

\newpage

\subsection{Additional sMIM data}
\begin{figure} [h]
\includegraphics[width =\linewidth]{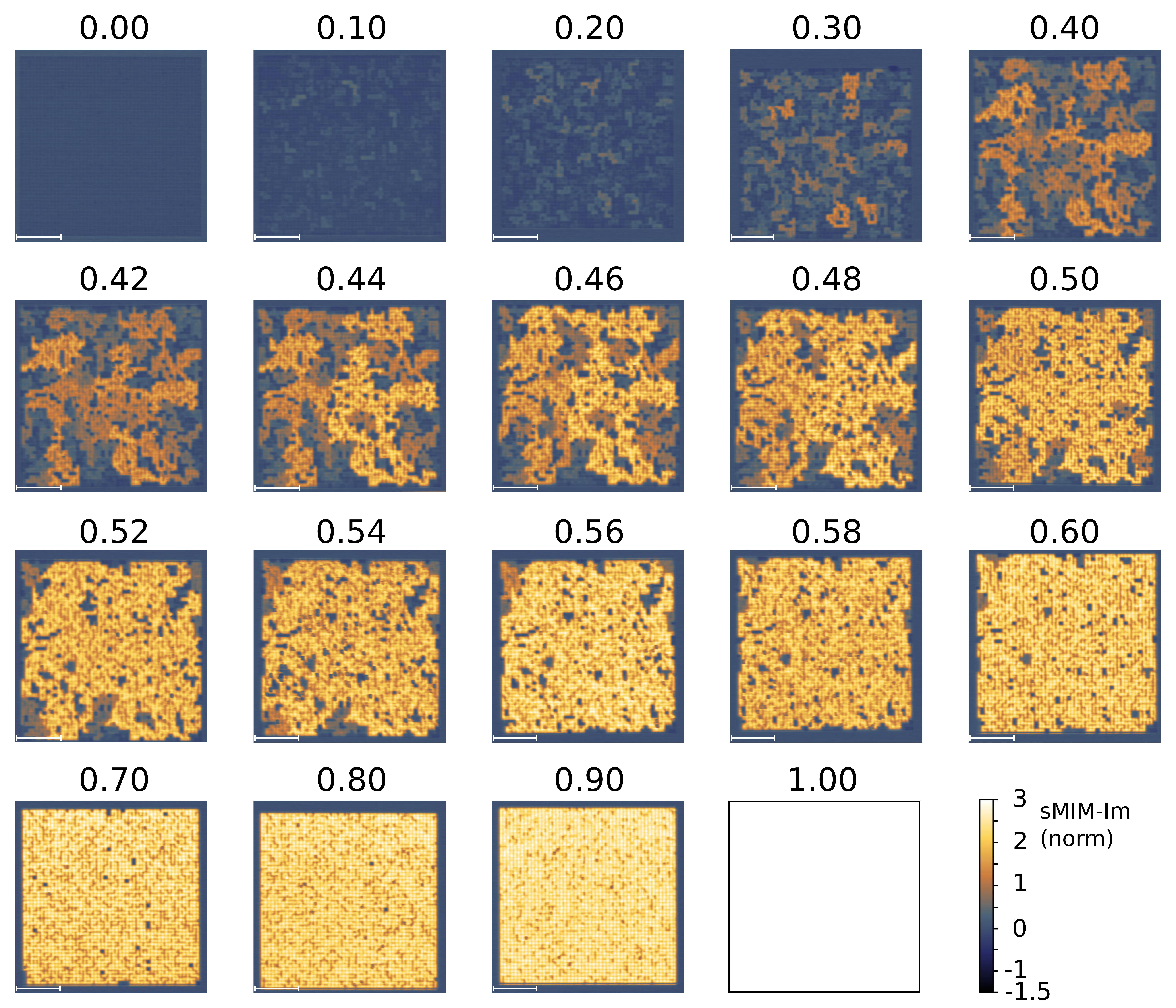}
\caption{\textbf{Aluminum- sMIM-Im: Experimental data}}
\end{figure}
\newpage
\begin{figure}[h]
\includegraphics[width = \linewidth]{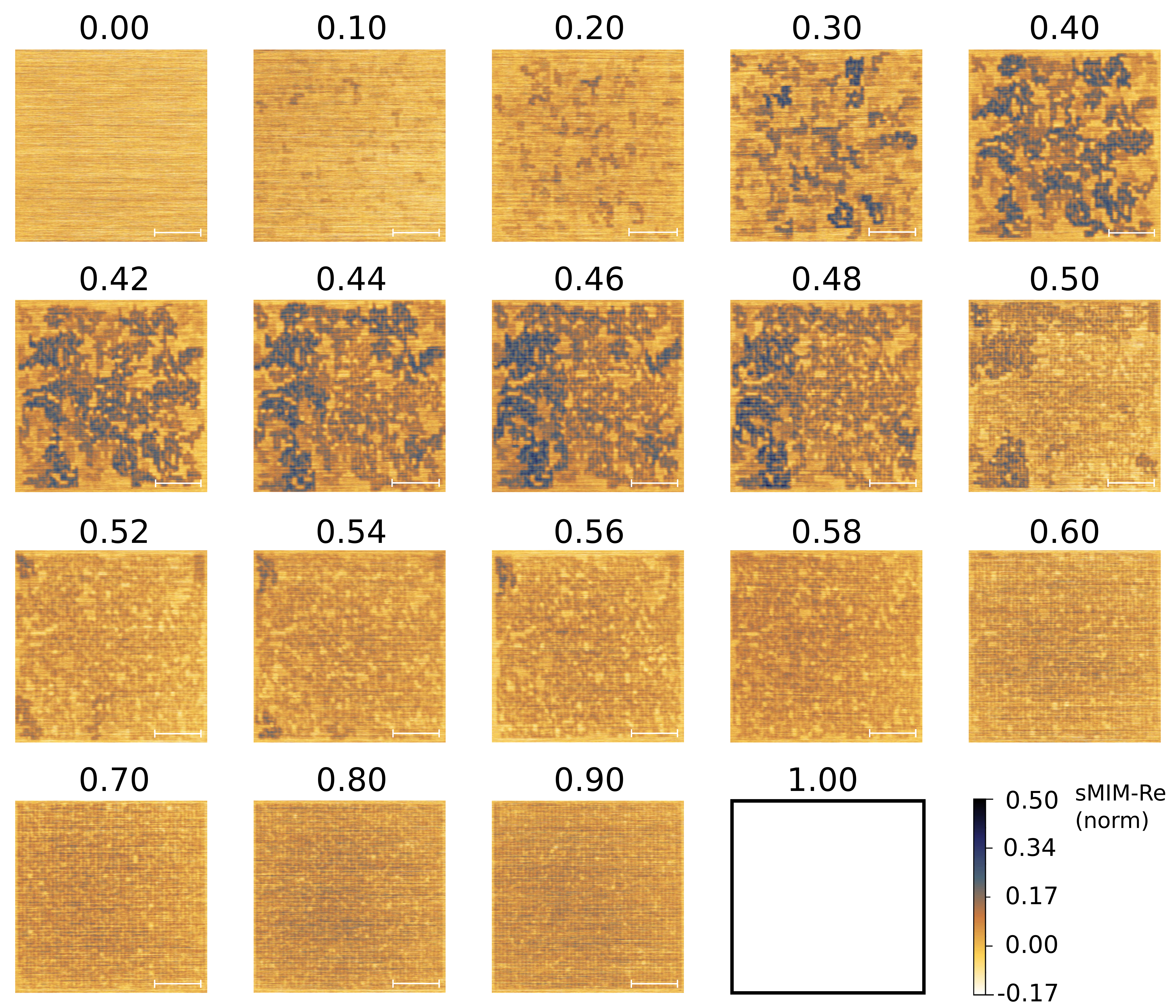}
\caption{\textbf{Aluminum - sMIM-Re: Experimental data}}
\end{figure}
\newpage
\begin{figure}[h]
\includegraphics[width = \linewidth]{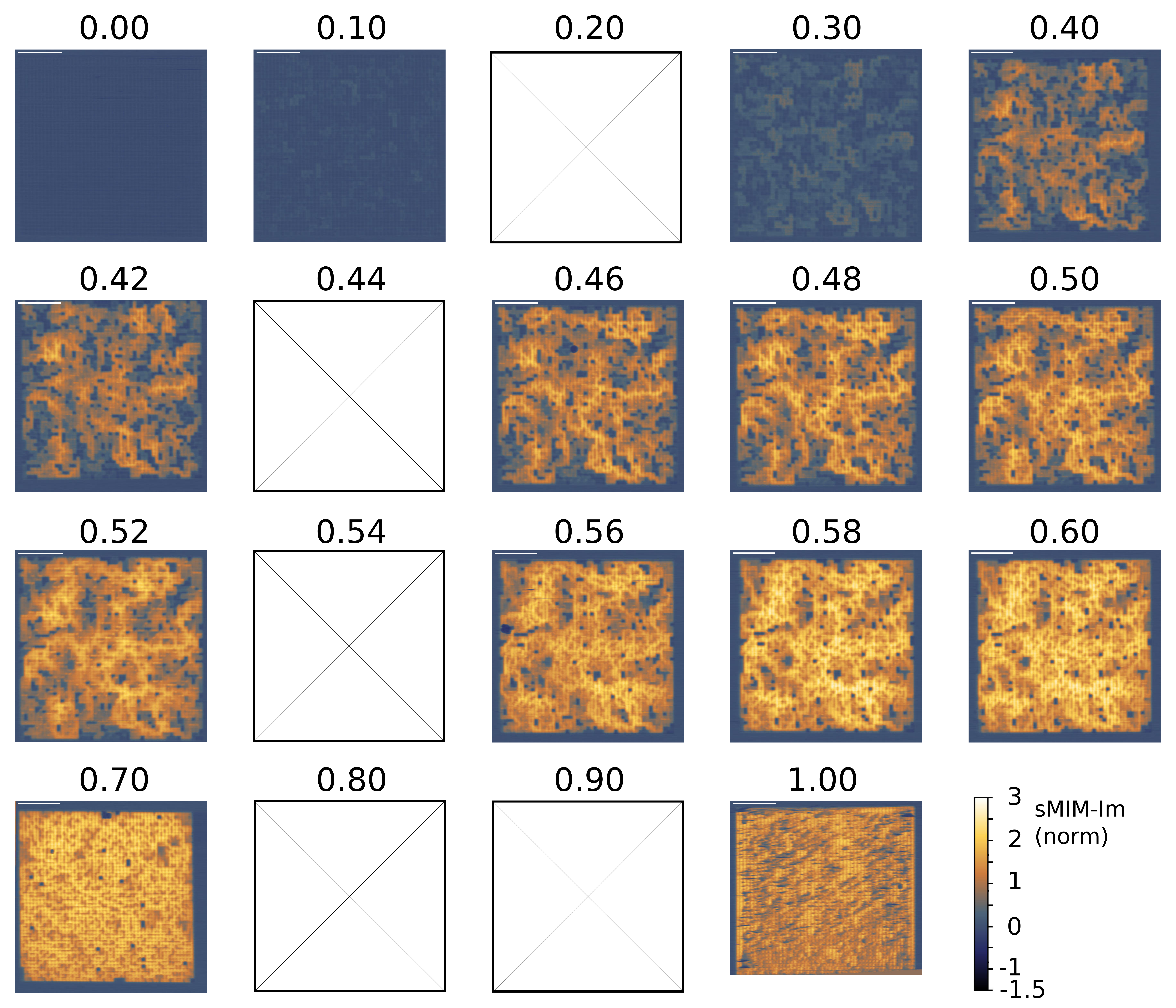}
\caption{\textbf{N\lowercase{b}T\lowercase{i}N - sMIM-I\lowercase{m}: Experimental data.} Fields marked with a cross could not be measured because the fabricated structures were damaged. For $p=1$, parts of the sample were covered with dirt residues from the sample fabrication, which appear as darker streaks in the image.}
\end{figure}
\newpage
\begin{figure}[h]
\includegraphics[width = \linewidth]{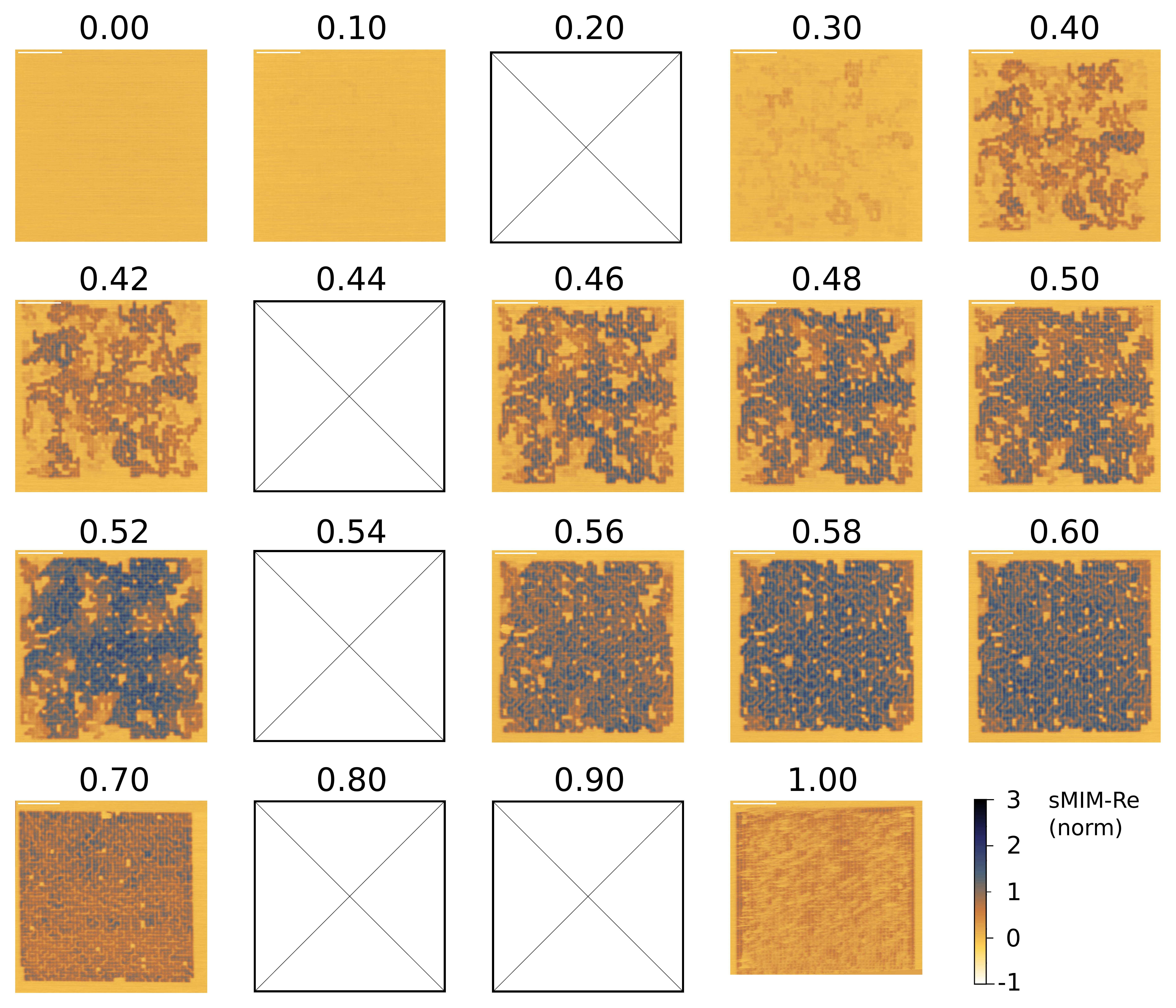}
\caption{\textbf{N\lowercase{b}T\lowercase{i}N - sMIM-R\lowercase{e}: Experimental data.} Fields marked with a cross could not be measured because the fabricated structures were damaged. For $p=1$, parts of the sample are covered with dirt residues from the sample fabrication. Here dissipation is reduced, leading to a suppressed sMIM-Re signal.}
\end{figure}

\newpage
\subsection{Results of the network model calculation}

\begin{figure} [h]
	\includegraphics[width = \linewidth]{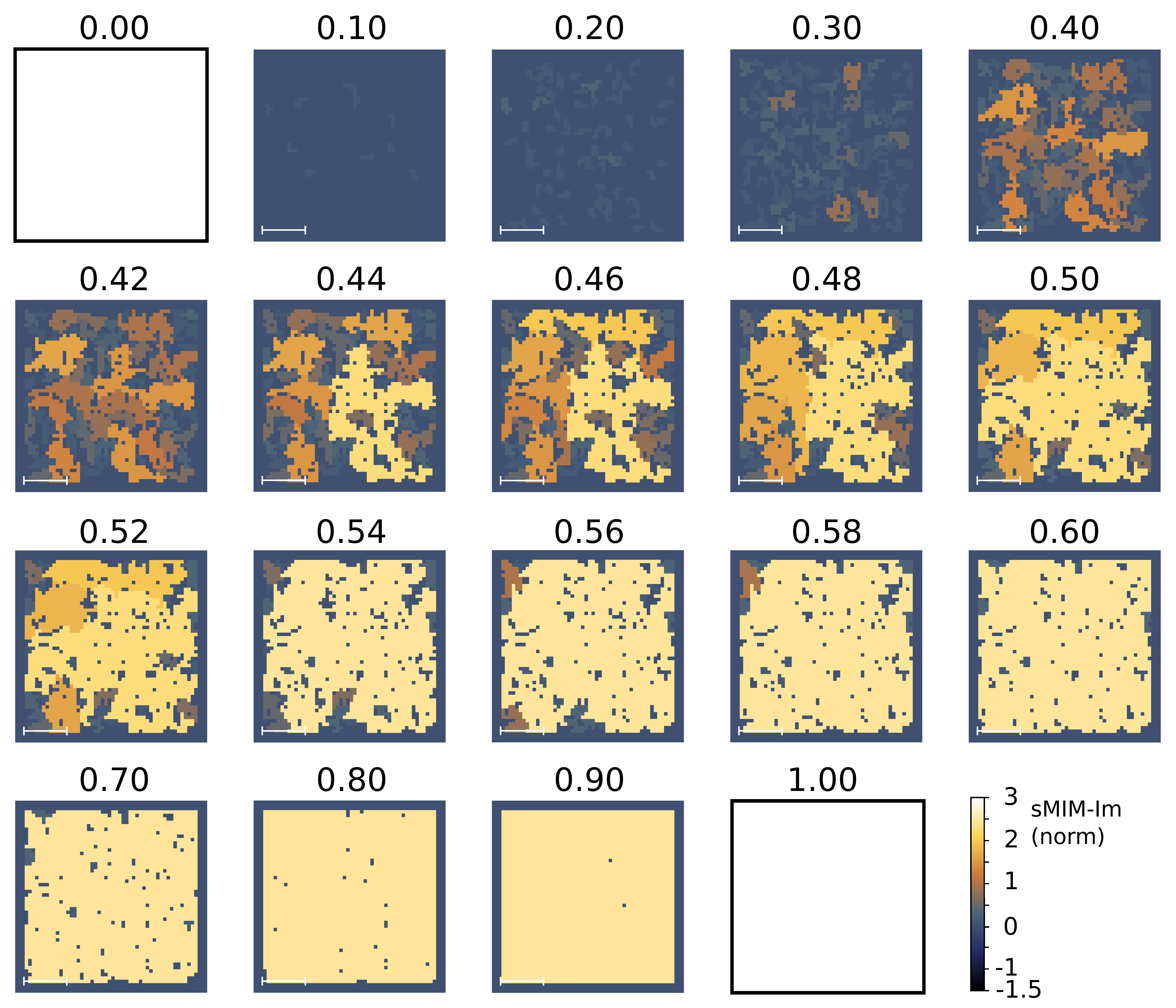}
	\caption{\textbf{Aluminum- sMIM-Im: Model calculations.}}
\end{figure}
\newpage
\begin{figure}[h]
	\includegraphics[width = \linewidth]{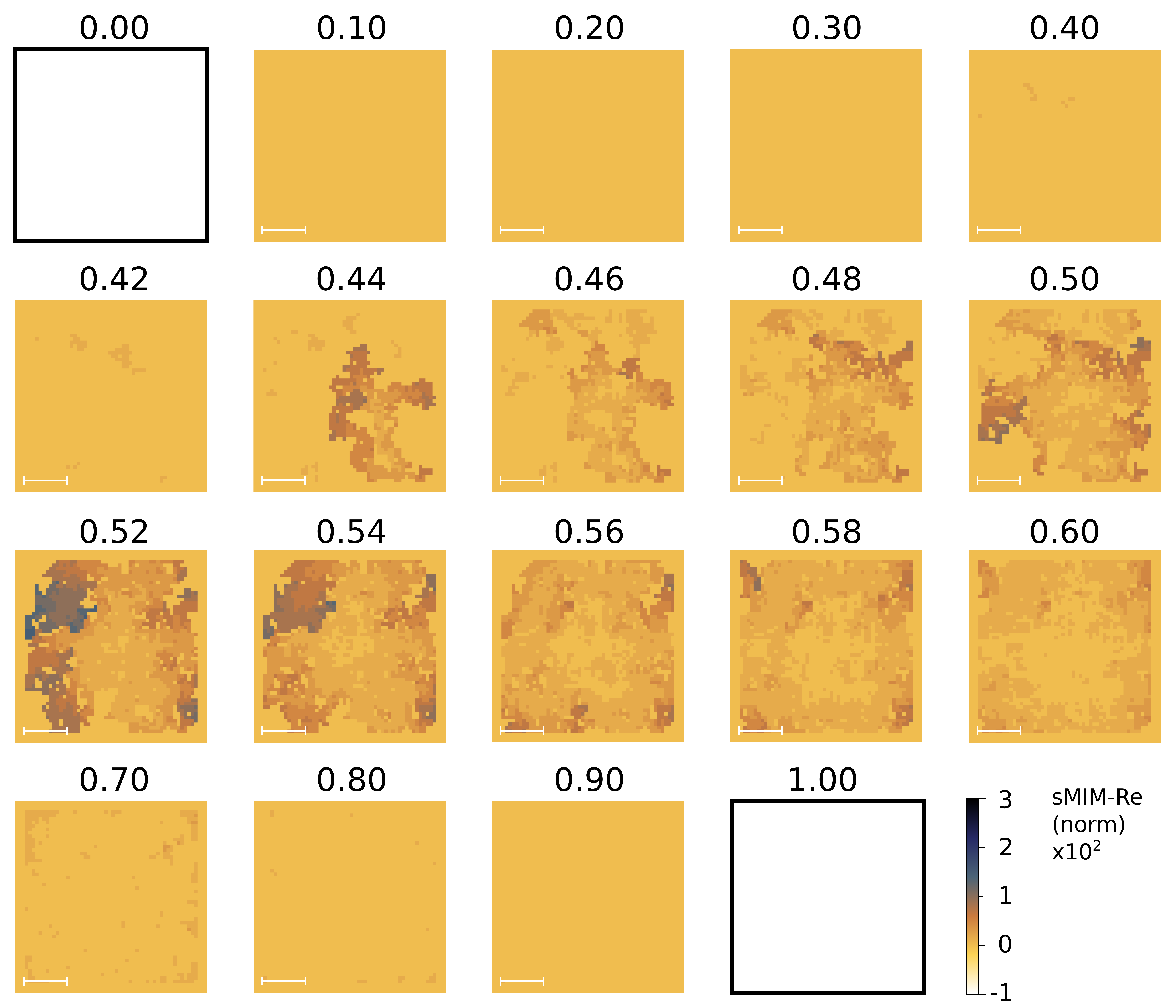}
	\caption{\textbf{Aluminum - sMIM-Re: Model calculations.}}
\end{figure}
\newpage
\begin{figure}[h]
	\includegraphics[width = \linewidth]{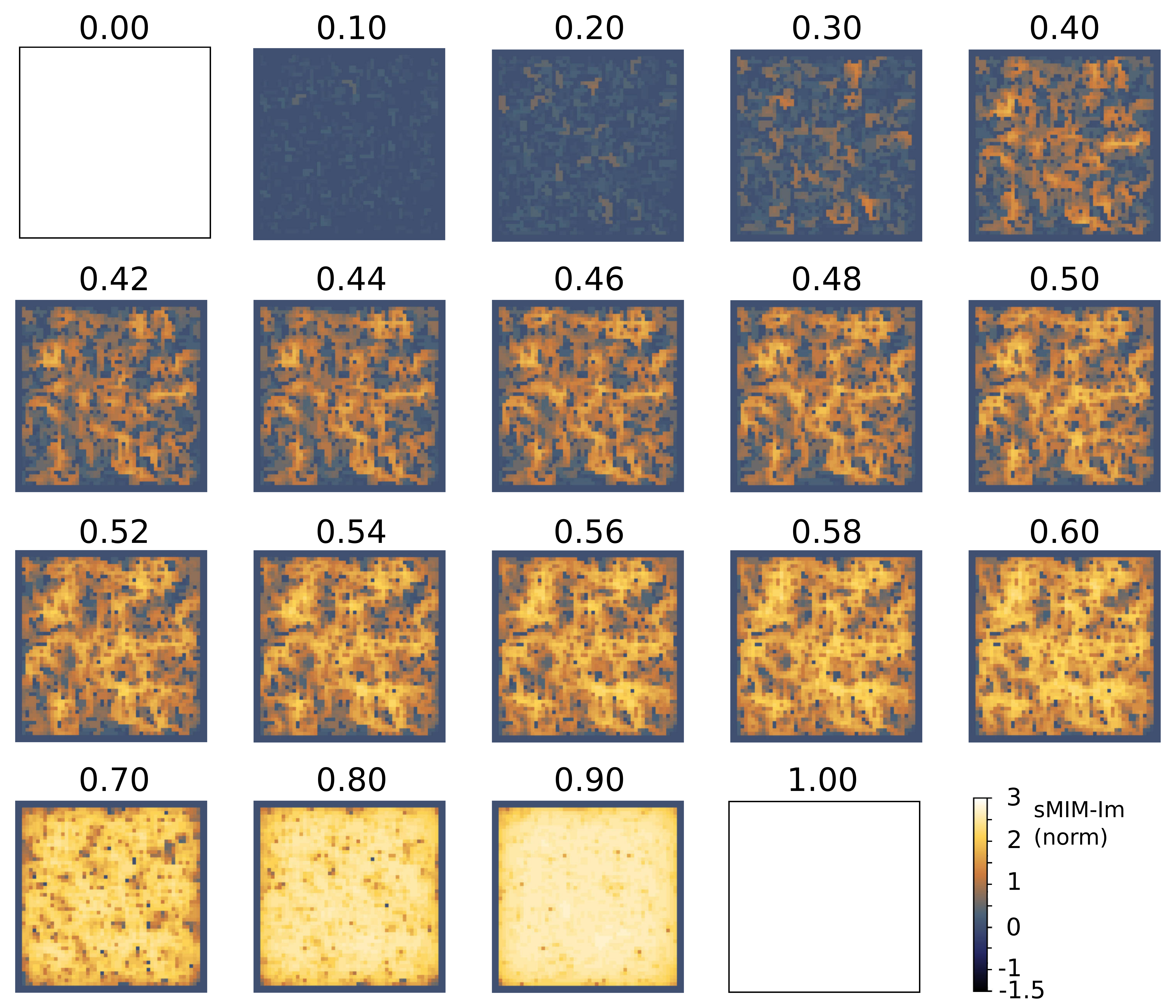}
	\caption{\textbf{N\lowercase{b}T\lowercase{i}N - sMIM-I\lowercase{m}: Model calculations.}}
\end{figure}
\newpage
\begin{figure}[h]
	\includegraphics[width = \linewidth]{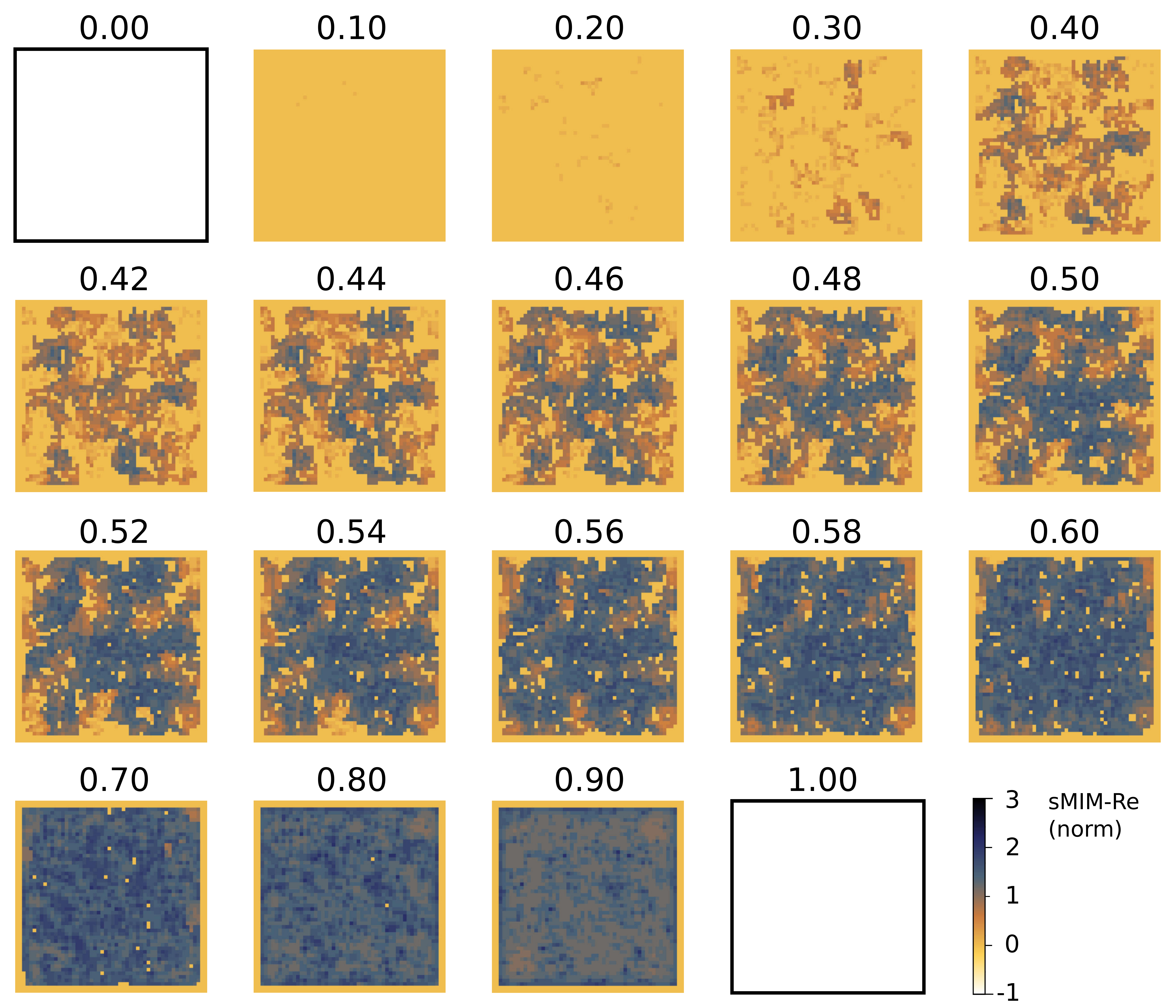}
	\caption{\textbf{N\lowercase{b}T\lowercase{i}N - sMIM-R\lowercase{e}: Model calculations.}}
\end{figure}

\end{document}